\providecommand{\Version}{3}

\ifnum\Version=1
	\def\Options{11pt,draftcls,onecolumn,twoside}
\fi
\ifnum\Version=2
	\def\Options{journal,twoside}
\fi
\ifnum\Version=3
	\def\Options{10pt,conference}
\fi
\ifnum\Version=4
	\def\Options{journal, onecoloumn}
\fi

\documentclass[\Options]{IEEEtran}

\ifnum\Version=4
	\usepackage[notref,notcite]{showkeys}
\fi

\usepackage{color}
\usepackage{bbm}
\usepackage{amssymb,amsmath,amsthm,mathrsfs}
\usepackage{dsfont}
\usepackage{graphicx}
\usepackage[hang]{subfigure}
\usepackage{mathtools}
\usepackage{bbm}
\usepackage{fixmath}

\usepackage{manfnt}

\theoremstyle{plain}
\newtheorem{thm}{Theorem}
\newtheorem{lem}{Lemma}

\theoremstyle{definition}
\newtheorem{defn}{Definition}

\theoremstyle{remark}
\newtheorem{rem}{Remark}

\DeclareMathOperator{\conv}{conv}

\newcommand{\rom}[1]{\uppercase\expandafter{\romannumeral #1\relax}}
\IEEEoverridecommandlockouts
\begin{document}

\title{Capacity Region Continuity of the Compound Broadcast Channel with Confidential Messages}
%\author{Andrea Grigorescu, Holger Boche, Rafael F. Schaefer and H. Vincent Poor
\author{
  \authorblockN{Andrea Grigorescu\IEEEauthorrefmark{1}, Holger Boche\IEEEauthorrefmark{1}, Rafael F. Schaefer\IEEEauthorrefmark{2} and H. Vincent Poor\IEEEauthorrefmark{2}\\[2mm]}
  \IEEEauthorblockA{ \IEEEauthorrefmark{1} Lehrstuhl f\"ur Theoretische Informationstechnik, Technische Universit\"at M\"unchen, 80333 M\"unchen, Germany}
  \authorblockA{\IEEEauthorrefmark{2} Department of Electrical Engineering, Princeton University, Princeton, NJ 08544, USA\\}
\thanks{This work of H. Boche was supported by the German Ministry of Education and Research (BMBF) under Grant 01BQ1050. This work of R. F. Schaefer was supported by the German Research Foundation (DFG) under Grant WY 151/2-1. This work of H. V. Poor was supported by the U.S. National Science Foundation under Grant CMMI-1435778.}
}

\maketitle
% ========================================================================================
% ========================================================================================
% ========================================================================================

\begin{abstract}
The compound \emph{broadcast channel with confidential messages} (BCC) generalizes the BCC by modeling the uncertainty of the channel. For the compound BCC, it is only known that the actual channel realization  belongs to a pre-specified uncertainty set of channels and that it is constant during the whole transmission.
For reliable and secure communication is necessary to operate at a rate pair within the compound BCC \emph{capacity region}. 
Therefor, the question whether small variations of the uncertainty set lead to large losses of the compound BCC capacity region is studied. It is shown that the compound BCC model is robust, i.e., the capacity region depends \emph{continuously} on the uncertainty set.
\end{abstract}

% ========================================================================================
% ========================================================================================
% ========================================================================================
\section{Introduction}
% === ITS
Information theoretic security was initiated by Wyner in \cite{wyner1975wire} introducing the wiretap channel, where the physical properties of the channel are used to guarantee security; see also \cite{liang2008information,bloch2011physical}. Subsequently, Csisz\'ar and K\"orner generalized the wiretap channel to the \emph{broadcast channel with confidential messages} (BCC) \cite{csiszar1978broadcast} using the \emph{weak secrecy} criterion.

% === CSI
For secure and reliable transmission over a wireless channel, channel state information (CSI) is needed, however, in practical systems it is not perfectly known.
\emph{Compound channels} model a simple and realistic CSI where the legitimate users are not aware of the actual channel realization. Nevertheless, they know it belongs to a known uncertainty set of channels and that it remains constant during the whole transmission. This model applies, for example, to the downlink of cellular system, where the base station transmits information to a user. The base station obtains limited CSI, for example via the uplink from pilot signal estimations at the receiver. Compound channels model the channel uncertainty based on a finite number of estimations.
\emph{Arbitrarily varying channels} model an even more limited CSI assumption. Here, it is assumed that the actual channel realization may additionally vary from channel use to channel use in an arbitrary fashion.

% === BCC
In this paper, the compound BCC is studied. 
The discrete memoryless compound BCC consists of one sender and two receivers. The sender wants to transmit two messages: a common message for both receivers and a confidential message for receiver 1. Receiver 2 must be kept ignorant from the confidential message.
In \cite{schaefer2014robust}, a multi-letter characterization of the compound BCC capacity region using the \emph{strong secrecy} criterion was established.

% === Continuity
In this work we investigate whether the capacity region of the compound BCC depends \emph{continuously} on the uncertainty set or not. If small changes of the uncertainty set cause large changes of the corresponding capacity region, the compound BCC is fragile, which complicates the design of practical communication systems. Hence, a continuous behavior of the capacity region is desired.

In \cite{boche2014continuity}, the continuity of the compound wiretap channel and arbitrarily varying wiretap channel (AVWC) was studied. The authors show that the secrecy capacity is continuous for the compound wiretap channel and discontinuous for the AVWC.

% === Contributions
Our main contribution is to show that the compound BCC capacity region depends continuously on the uncertainty set. Using a channel example from \cite{boche2014continuity}, we state that the capacity region of the arbitrarily varying BCC (AVBCC) is discontinuous, which shows continuity of the compound BCC capacity region cannot be generalized to the AVBCC.

% === Outline
In Section \ref{sec2} we introduce the compound BCC and its capacity region. In Section \ref{sec3} we introduce a distance between two compound BCC and a distance between two sets and we show that the capacity region of the compound BCC is a continuous function of the uncertainty set. Finally, we conclude our paper with a discussion in Section \ref{sec4}. 
% === Notation
\footnote{\emph{Notation}: $\mathbb{N}$ and $\mathbb{R}_+$ denote the sets of non-negative integers and non-negative real numbers, respectively; $\mathcal{I}=(\cdot,\cdot)$ and $\mathcal{J}=[\cdot,\cdot]$ denote open and closed interval, respectively; $\overline{\conv}(\mathcal{A})$ denotes the convex hull closure of the set $\mathcal{A}$;  $H(\cdot)$, $H_{2}(\cdot)$, $I(\cdot ;\cdot)$ are the entropy, binary entropy, and mutual information,respectively; all logarithms and information quantities are taken to the base $2$; $\|\nu-\mu\|\coloneqq\sum_{a\in\mathcal{A}}|\nu(a)-\mu(a)|$ is the total variation distance of measures $\mu$ and $\nu$ on $\mathcal{A}$; the space of probability distribution on the finite set $\mathcal{A}$ is denoted by $\mathcal{P}(\mathcal{A})$.}

% ========================================================================================
% ========================================================================================
% ========================================================================================

\section{Compound Broadcast Channel with Confidential Messages}
\label{sec2}

The transmitter and the receiver of a compound channel know an uncertainty set of channels to which the channel belongs, however, they do not know the actual channel realization. The channel remains constant during the whole transmission. We consider a two receiver compound BCC. The transmitter sends simultaneously a common message to both receivers and a confidential message to receiver 1, which must be kept secret from receiver 2.
Let $\mathcal{X}$ be the finite input alphabet, $\mathcal{Y}$ and $\mathcal{Z}$ the finite output alphabets of receivers 1 and 2, respectively, and let $\mathcal{S}$ be a finite set of channel states. For each channel state $s\in\mathcal{S}$, input sequence $x^n\in\mathcal{X}^n$ and output sequences $y^n\in\mathcal{Y}^n$ and $z^n\in\mathcal{Z}^n$, the discrete memoryless broadcast channel is given by $Q_s^n(y^n,z^n| x^n)\coloneqq\prod_{i=1}^n Q_s(y_i,z_i| x_i)$ with marginal channels $W^n_{s}(y^n| x^n)$ and $V^n_{s}(z^n| x^n)$.
\begin{defn}
The discrete memoryless \emph{compound broadcast channel} $\mathfrak{W}$ is given by the channel pair family with common input

\begin{equation*}
\mathfrak{W}\coloneqq \{(W_{s},V_{s})\colon s \in \mathcal{S}\}.
\end{equation*}
\end{defn}
% === Codes for BCC

\subsection{Codes for Compound Broadcast Channels}

We consider a block-code of arbitrary but fixed length $n$. Let $\mathcal{M}_0:=\{1,\dotsc,M_{0,n}\}$ be the common message set and $\mathcal{M}_1:=\{1,\dotsc,M_{1,n}\}$ the confidential message set. We use the abbreviation $\mathcal{M}\coloneqq \mathcal{M}_0 \times \mathcal{M}_1$.
\begin{defn}
An $(n,M_{0,n},M_{1,n})$-\emph{code} for the compound BCC consists of a stochastic encoder
\begin{equation*}
E\colon \mathcal{M}_0 \times \mathcal{M}_1 \rightarrow \mathcal{P}(\mathcal{X}^n)
\end{equation*}
i.e., a stochastic matrix, and decoders at receivers $1$ and $2$
\begin{align*}
\varphi_1 &\colon \mathcal{Y}^n\rightarrow\mathcal{M}_0 \times \mathcal{M}_1\\
\varphi_2 &\colon \mathcal{Z}^n\rightarrow\mathcal{M}_0. 
\end{align*} 
\end{defn}

The average error probability for receivers 1 and 2 and the channel realization $s\in\mathcal{S}$ are
\begin{align*}
\overline{e}_{1,n}(s)&:=\frac{1}{|\mathcal{M}|}\sum_{m\in\mathcal{M}}\sum_{x^n\in\mathcal{X}^n}\sum_{y^n:\varphi_1(y^n)\neq m}\!\!\!\!\!\!W_s(y^n|x^n)E(x^n|m)\\
\overline{e}_{2,n}(s)&:=\frac{1}{|\mathcal{M}|}\sum_{m\in\mathcal{M}}\sum_{x^n\in\mathcal{X}^n}\sum_{z^n:\varphi_2(z^n)\neq m_0}\!\!\!\!\!\!V_s(z^n|x^n)E(x^n|m).
\end{align*}
Since reliable communication is required for all $s \in\mathcal{S}$, we consider the maximum average error probabilities, i.e. $\overline{e}_{1,n}=\max_{s\in\mathcal{S}}\overline{e}_{1,n}(s)$ and $\overline{e}_{2,n}=\max_{s\in\mathcal{S}}\overline{e}_{2,n}(s)$.

The confidential message has to be kept secret from the non-legitimate receiver for all channel realizations. Therefore, we require $\max_{s\in\mathcal{S}}I(M_1;Z_s^n)\leq\epsilon_n$ for some $\epsilon_n>0$ with $M_1$ the random variable uniformly distributed over the set $\mathcal{M}_1$ and $Z_s^n$ the output at the non-legitimate receiver for the channel realization $s\in\mathcal{S}$. This criterion is known as \emph{strong secrecy} \cite{csiszar1996almost,maurer2000information}.
\begin{defn}
A rate pair $(R_0,R_1)\in \mathbb{R}^2_{+}$ is said to be achievable for the compound BCC if for any $\tau>0$ there is an $n(\tau)\in \mathbb{N}$ and a sequence of $(n,M_{0,n},M_{1,n})$-codes such that for all $n\geq n(\tau)$ we have $\frac{1}{n}\log M_{0,n}\geq R_0-\tau$,$\frac{1}{n}\log M_{1,n}\geq R_1-\tau$, and 
\begin{align}
\max_{s\in \mathcal{S}}I(M_1;Z_s^n)\leq \epsilon_n
\end{align}
with $\overline{e}_{1,n},\overline{e}_{2,n},\epsilon_n\rightarrow0$ as $n\rightarrow \infty$.
\end{defn}
\begin{defn}
The set closure of all achievable rate pairs is the \emph{capacity region} $\mathcal{C}(\mathfrak{W})$ of the compound BCC $\mathfrak{W}$.
\end{defn}

\subsection{Capacity Results}
In this section we present an achievable rate region and a multi-letter characterization of the compound BCC capacity region \cite{schaefer2014robust}.  
\begin{lem} [{\cite{schaefer2014robust}}]
An achievable secrecy rate region for the compound BCC is given by the set of all rate pairs $(R_0,R_1) \in \mathbb{R}^2_+$ satisfying
\begin{align*}
R_0 &\leq \min_{s\in\mathcal{S}}\min\{I(U;Y_s),I(U;Z_s)\}\\
R_1 &\leq \min_{s\in\mathcal{S}}I(V;Y_s| U) - \max_{s\in\mathcal{S}}I(V;Z_s| U)
\end{align*}
for some random variables $U,V,X$ where $U-V-X-(Y_s,Z_s)$ forms a Markov chain. 
Furthermore, the strong secrecy criterion goes exponentially fast to zero and the decoding error at the non-legitimate receiver goes exponentially fast to one.
\end{lem}
We next present a multi-letter description of $\mathcal{C}(\mathfrak{W})$ of the compound BCC $\mathfrak{W}$. Let $n\in\mathbb{N}$ be arbitrary but fixed. We define the rate region $\mathcal{R}_n(\mathfrak{W},U,V,X^n)$ as the set of all rate pairs $(R_0,R_1)\in \mathbb{R}^2_+$ satisfying
\begin{align}
R_0 &\leq \frac{1}{n} \inf_{s\in\mathcal{S}}\min\{I(U;Y_s^n),I(U;Z_s^n)\}\label{lim_R0}\\
R_1 &\leq  \frac{1}{n}(\inf_{s\in\mathcal{S}}I(V;Y_s^n| U) - \sup_{s\in\mathcal{S}}I(V;Z_s^n| U))\label{lim_R1}
\end{align}
for the random variables satisfying the Markov chain relationship $U-V-X^n-(Y^n_s,Z^n_s)$. 
For a given $n\in\mathbb{N}$ we define the region 
\begin{equation*}
\mathcal{M}_n(\mathfrak{W})= \bigcup_{U-V-X^n}\mathcal{R}_n(\mathfrak{W},U,V,X^n)
\end{equation*}
that is, $\mathcal{M}_n(\mathfrak{W})$ is the union of the regions $\mathcal{R}_n(\mathfrak{W},U,V,X^n)$ over all random variables satisfying the Markov chain relationship $U-V-X^n$.
\begin{thm}
The strong secrecy capacity region $\mathcal{C}(\mathfrak{W})$ of the compound BCC $\mathfrak{W}$ is the  convex hull closure of the union of the regions $\mathcal{M}_n(\mathfrak{W})$ over all $n\in\mathbb{N}$, i.e.
\begin{equation*}
\mathcal{C}(\mathfrak{W})=\overline{\conv}({\bigcup_{n\in\mathbb{N}}\mathcal{M}_n(\mathfrak{W})}).
\end{equation*}
\end{thm}
\begin{rem}
To the best of our knowledge, there is still no single-letter characterization of $\mathcal{C}(\mathfrak{W})$ known.
\end{rem}
\begin{rem}
The union of the rate regions $\bigcup_{n\in\mathbb{N}}\mathcal{M}_n(\mathfrak{W})$ may itself not be convex. However, all rate pairs in the convex hull can be achieved by time sharing between the points in the rate regions $\mathcal{M}_n(\mathfrak{W})$. 
\end{rem}
%\begin{rem}
%The strong secrecy criterion goes exponentially fast to zero and the decoding error at the non-legitimate receiver goes exponentially fast to one. This is achieved by compound BCC codes satisfying the vanishing output variation showed in \cite{schaefer2014robust}.
%\end{rem}

% ========================================================================================
% ========================================================================================
% ========================================================================================

\section{Continuity of the Compound BCC Capacity Region}
\label{sec3}
In this section we first define the distance between two compound BCCs and the distance between rate regions. We then analyze the continuity of the compound BCC capacity region. 
% ========================================================================================
\subsection{Distance between Compound Broadcast Channels and Sets}
Let $(W,V)$ and $(\widetilde{W},\widetilde{V})$ be two broadcast channels. We define the distance between channels as
\begin{align*}
d(W,\widetilde{W}) & \coloneqq \max_{x\in\mathcal{X}}\sum_{y\in\mathcal{Y}}| W(y|x)-\widetilde{W}(y|x)|\\
d(V,\widetilde{V}) & \coloneqq \max_{x\in\mathcal{X}}\sum_{z\in\mathcal{Z}}| V(z|x)-\widetilde{V}(z|x)|
\end{align*}
and the distance between two broadcast channels as
\begin{equation*}
d((W,V),(\widetilde{W},\widetilde{V})) \coloneqq \max(d(W,\widetilde{W}),d(V,\widetilde{V})).
\end{equation*} 

Let $\mathfrak{W}_1=\{(W_{s_1},V_{s_1})\colon s_1\in \mathcal{S}_1\}$ and $\mathfrak{W}_2=\{(W_{s_2},V_{s_2})\colon s_2\in \mathcal{S}_2\}$ be two finite compound broadcast channels with marginal compound channels $\mathcal{W}_i=\{W_{s_i}\colon s_i\in \mathcal{S}_i\}$ and $\mathcal{V}_i=\{V_{s_i}\colon s_i\in \mathcal{S}_i\}$ for $i\in\{1,2\}$. We define the distance between two marginal compound channels as
\begin{align*}
d_1(\mathcal{W}_1,\mathcal{W}_2) &= \max_{s_2\in\mathcal{S}_2}\min_{s_1\in\mathcal{S}_1}d(W_{s_1},W_{s_2})\\
d_2(\mathcal{W}_1,\mathcal{W}_2) &= \max_{s_1\in\mathcal{S}_1}\min_{s_2\in\mathcal{S}_2}d(W_{s_1},W_{s_2})
\end{align*}
\begin{align*}
d_1(\mathcal{V}_1,\mathcal{V}_2) &= \max_{s_2\in\mathcal{S}_2}\min_{s_1\in\mathcal{S}_1}d(V_{s_1},V_{s_2})\\
d_2(\mathcal{V}_1,\mathcal{V}_2) &= \max_{s_1\in\mathcal{S}_1}\min_{s_2\in\mathcal{S}_2}d(V_{s_1},V_{s_2}).
\end{align*}
\begin{defn}
Let $\mathfrak{W}_1$ and $\mathfrak{W}_2$ be two compound broadcast channels. The distance $D(\mathfrak{W}_1,\mathfrak{W}_2)$ between $\mathfrak{W}_1$ and $\mathfrak{W}_2$ is defined as
\begin{align*}
D(\mathfrak{W}_1,\mathfrak{W}_2)= \max \Big\{&d_1(\mathcal{W}_1,\mathcal{W}_2),d_2(\mathcal{W}_1,\mathcal{W}_2),\\
&d_1(\mathcal{V}_1,\mathcal{V}_2),d_2(\mathcal{V}_1,\mathcal{V}_2)\Big\}.
\end{align*}
\end{defn}
%Let $\mathfrak{W}$ be a compound broadcast channel with state set $\mathcal{S}$ and further let $\{\mathfrak{W}_n\}_{n\in\mathbb{N}}$ be a sequences with state set sequence $\{\mathcal{S}_n\}_{n\in\mathbb{N}}$ satisfying
%\begin{equation*}
%\lim_{n\rightarrow\infty}D(\mathfrak{W},\mathfrak{W}_n)=0
%\end{equation*}
%Then for every channel realization $(W_{\hat{s}},V_{\hat{s}})\in\{(W_s,V_s)\}_{s\in\mathcal{S}}$ there exists a sequence $\{\hat{s}_n\}_{n\in\mathbb{N}}$ with $\hat{s}_n\in\mathcal{S}_n$ and
%\begin{align*}
%\lim_{n\rightarrow\infty}\max_{x\in\mathcal{X}}\sum_{y\in\mathcal{Y}}\Big|W_{\hat{s}}(y|x)-W_{\hat{s}_n}(y|x)\Big|&=0\\
%\lim_{n\rightarrow\infty}\max_{x\in\mathcal{X}}\sum_{z\in\mathcal{Z}}\Big|V_{\hat{s}}(z|x)-V_{\hat{s}_n}(z|x)\Big|&=0.
%\end{align*}
%Moreover, for every $\epsilon>0$ there is a $n_0=n_0(\epsilon)$ such that for all $n\geq n_0$ and all $(W_{\hat{s}},V_{\hat{s}})\in\mathfrak{W}_n$ there is a channel realization $(W_{\tilde{s}},V_{\tilde{s}})\in\mathfrak{W}$ such that
%\begin{align*}
%\max_{x\in\mathcal{X}}\sum_{y\in\mathcal{Y}}\Big|W_{\hat{s}}(y|x)-W_{\tilde{s}}(y|x)\Big|&<\epsilon\\
%\max_{x\in\mathcal{X}}\sum_{z\in\mathcal{Z}}\Big|V_{\hat{s}}(z|x)-V_{\tilde{s}}(z|x)\Big|&<\epsilon
%\end{align*}

To compare different rate regions, we define the following distance of sets.
\begin{defn}
Let $\mathcal{R}_1,$ and $\mathcal{R}_2$ be two non-empty compact subsets of the metric space $(\mathbb{R}^2_+,d)$ with $d(x,y)={\sum_{i=1}| x_i-y_i|}$ for all $x,y \in \mathbb{R}$. We define the distance between two sets as
\begin{align*}
D_R(\mathcal{R}_1,\mathcal{R}_2)=\max\big\{&\max_{r_1\in\mathcal{R}_1}\min_{r_2\in\mathcal{R}_2}d(r_1,r_2),\\
&\max_{r_2\in\mathcal{R}_2}\min_{r_1\in\mathcal{R}_2}d(r_1,r_2)\big\}.
\end{align*}
\end{defn}

% ========================================================================================

\subsection{Continuity of the Capacity Region of the Compound BCC}
We use the following technical result, which is an extension of Lemma 2 from \cite{boche2014continuity}.
\begin{lem}[{\cite{boche2014continuity}}]\label{lem1}
Let $\epsilon \in (0,1)$ be arbitrary. For all $(X,Y)$ and $(\widetilde{X},\widetilde{Y})$ be two pairs of random variables with finite range $\mathcal{X}\times\mathcal{Y}$ and joint probabilities distributions $P_{X,Y},P_{\widetilde{X},\widetilde{Y}}\in\mathcal{P}(\mathcal{X}\times\mathcal{Y})$. If $|| P_{X,Y}-P_{\widetilde{X},\widetilde{Y}}||\leq\epsilon$, then it holds
\begin{equation}\label{delta1}
| H(Y| X)-H(\widetilde{Y}|\widetilde{X})|\leq\delta_1(\epsilon,|\mathcal{Y}|)
\end{equation}
with $\delta_1(\epsilon,|\mathcal{Y}|):=2\epsilon\log|\mathcal{Y}|+2H_2(\epsilon)$.
\end{lem}
%\begin{rem}
%Note that the Big hand side of \eqref{delta1} depends only on the size of the alphabet $\mathcal{Y}$, but it is independent of $\mathcal{X}$. 
%\end{rem}
\begin{lem}\label{lem2}
Let $\mathcal{X}$ and $\mathcal{Y}$ be finite alphabets and $W,\widetilde{W}\colon \mathcal{X}\rightarrow\mathcal{P}(\mathcal{Y})$ be arbitrary channels with
\begin{equation*}
d(W,\widetilde{W})\leq \epsilon
\end{equation*}
for some $\epsilon>0$. For an arbitrary $n \in \mathbb{N}$, let $\mathcal{U}$ and $\mathcal{V}$ be two finite sets, $P_U\in\mathcal{P}(\mathcal{U})$ the uniform distribution on $\mathcal{U}$, $P_{V|U}(\cdot|u)$ is the conditional distribution of the random variable $V$ over $\mathcal{V}$ given $U=u$ and $E(x^n| u)$ with $x^n \in \mathcal{X}^n$ conditioned on $u\in\mathcal{U}$ is an arbitrary stochastic encoder. We consider the probability distributions
\begin{align*}
P_{UVY^n}(u,v,y^n)=\!\!\!\!\sum_{x^n\in\mathcal{X}^n}\!\!\!W^n(y^n| x^n)E(x^n| v)P_{V|U}(v|u)P_U(u)\\
P_{UV\widetilde{Y}^n}(u,v,\tilde{y}^n)=\!\!\!\!\sum_{x^n\in\mathcal{X}^n}\!\!\!\widetilde{W}^n(y^n| x^n)E(x^n|v)P_{V|U}(v|u)P_U(u)
\end{align*}
Then it holds
\begin{equation}\label{delta2}
| I(V;Y^n|U)-I(V;\widetilde{Y}^n|U)|\leq n\delta_2(\epsilon,|\mathcal{Y}|)
\end{equation}
with $\delta_2(\epsilon,|\mathcal{Y}|):=4\epsilon\log|\mathcal{Y}|+4H_2(\epsilon)$.
\end{lem}
\begin{IEEEproof}
See the arxiv version of this work \cite{grigorescu2014continuity}.
\end{IEEEproof}
\begin{rem}
Note that the right-hand side of \eqref{delta2} and \eqref{delta1} depend only on the size of the output alphabet $\mathcal{Y}$, but they are independent of the size of the auxiliary alphabets $\mathcal{U}$ and $\mathcal{V}$, the conditional distribution $P_{V|U}$ and the chosen stochastic encoder $E$. 
\end{rem}
\begin{lem}\label{ratecont}
Let $\epsilon\in(0,1)$ and $n\in\mathbb{N}$. Let $\mathfrak{W}_1$ and $\mathfrak{W}_2$ be two compound BCCs and random variables satisfying the Markov chain relationship $U-V-X^n$. If
\begin{equation*}
D(\mathfrak{W}_1,\mathfrak{W}_2)\leq\epsilon
\end{equation*}
then it holds
\begin{equation*}
D_R({\mathcal{R}_n(\mathfrak{W}_1,U,V,X^n)},{\mathcal{R}_n(\mathfrak{W}_2,U,V,X^n)})\leq \delta(\epsilon,|\mathcal{Y}|,|\mathcal{Z}|)
\end{equation*}
with $\delta(\epsilon,|\mathcal{Y}|,|\mathcal{Z}|)= {\delta'(\epsilon,|\mathcal{Y}|,|\mathcal{Z}|)+\delta''(\epsilon,|\mathcal{Y}|,|\mathcal{Z}|)}$, $\delta'(\epsilon,|\mathcal{Y}|,|\mathcal{Z}|)\coloneqq 4H_2(\epsilon)+4\epsilon\max\{\log|\mathcal{Y}|,\log|\mathcal{Z}|\}$ and $\delta''(\epsilon,|\mathcal{Y}|,|\mathcal{Z}|)\coloneqq4\epsilon\log|\mathcal{Y}||\mathcal{Z}|+8H_2(\epsilon)$.
\end{lem}
\begin{IEEEproof}
%Let $\mathcal{R}_n(\mathfrak{W}_1,U,V,X^n)$ and $\mathcal{R}_n(\mathfrak{W}_2,U,V,X^n)$ be the achievable rate regions for two compound BCCs. 
The regions $\mathcal{R}_n(\mathfrak{W}_1,U,V,X^n)\in\mathbb{R}^2_+$ and $\mathcal{R}_n(\mathfrak{W}_2,U,V,X^n)\in\mathbb{R}^2_+$ are rectangles described by the rates $(R_{0,\mathcal{S}_1},R_{1,\mathcal{S}_1})$ and $(R_{0,\mathcal{S}_2},R_{1,\mathcal{S}_2})$ satisfying (\ref{lim_R0}) and (\ref{lim_R1}) respectively.
%\begin{align*}
%\mathcal{R}_n(\mathfrak{W}_1,U,V,X^n)=
%\Big\{(R_{0,\mathcal{S}_1},R_{1,\mathcal{S}_1})\colon
%\begin{array}{l}
%0\leq R_{0,\mathcal{S}_1} \leq \frac{1}{n}\inf_{s_1\in \mathcal{S}_1}\min\{I(U;Y^n_{s_1}),I(U;Z^n_{s_1})\}\\
%0\leq R_{1,\mathcal{S}_1} \leq \frac{1}{n}\inf_{s_1\in \mathcal{S}_1}I(V;Y^n_{s_1}| U)-\frac{1}{n}\sup_{s_1\in %\mathcal{S}_1}I(V;Z^n_{s_1}| U)
%\end{array}
%\Big\}
%\end{align*}
%and 
%\begin{align*}
%\mathcal{R}_n(\mathfrak{W}_2,U,V,X^n)=
%\Big\{(R_{0,\mathcal{S}_2},R_{1,\mathcal{S}_2})\colon
%\begin{array}{l}
%0\leq R_{0,\mathcal{S}_2} \leq \frac{1}{n}\inf_{s_2\in \mathcal{S}_2}\min\{I(U;Y^n_{s_2}),I(U;Z^n_{s_2})\}\\
%0\leq R_{1,\mathcal{S}_2} \leq \frac{1}{n}\inf_{s_2\in \mathcal{S}_2}I(V;Y^n_{s_2}| U)-\frac{1}{n}\sup_{s_2\in \mathcal{S}_2}I(V;Z^n_{s_2}| U)
%\end{array}
%\Big\}
%\end{align*}
For $i=1,2$, we define $A_{0_{\mathcal{S}_i}}$ and $A_{1_{\mathcal{S}_i}}$
\begin{align*}
A_{0_{\mathcal{S}_i}}&= \max_{(R_{0,\mathcal{S}_i},R_{1,\mathcal{S}_i})\in\mathcal{R}_n(\mathfrak{W}_i,U,V,X^n)} R_{0,\mathcal{S}_i}\\ 
A_{1_{\mathcal{S}_i}}&= \max_{(R_{0,\mathcal{S}_i},R_{1,\mathcal{S}_i})\in\mathcal{R}_n(\mathfrak{W}_i,U,V,X^n)} R_{1,\mathcal{S}_i}.
\end{align*} 
Note that both regions are rectangles sharing the corner point $(0,0)$. Therefore, the longest distance between these two sets is given by the corner points $(A_{0_{\mathcal{S}_1}},A_{1_{\mathcal{S}_1}})$ and $(A_{0_{\mathcal{S}_2}},A_{1_{\mathcal{S}_2}})$, i.e.,
\begin{align*}
D_R(&\mathcal{R}_n(\mathfrak{W}_1,U,V,X^n),\mathcal{R}_n(\mathfrak{W}_2,U,V,X^n))\\
&={| A_{0_{\mathcal{S}_1}}-A_{0_{\mathcal{S}_2}}|+| A_{1_{\mathcal{S}_1}}- A_{1_{\mathcal{S}_2}}|}.
\end{align*}
We first analyze the difference between the maximum achievable common rates, i.e., $| A_{0_{\mathcal{S}_1}}-A_{0_{\mathcal{S}_2}}|$ and then the difference between the maximum achievable confidential rates, i.e., $| A_{1_{\mathcal{S}_1}}- A_{1_{\mathcal{S}_2}}|$. 
\subsubsection{Common Message Rate}
There are four cases that may occur: 
\begin{enumerate}
\renewcommand{\labelenumi}{\textbf{\theenumi}}
\renewcommand{\theenumi}{\arabic{enumi})}
\item $A_{0_{\mathcal{S}_1}}=\frac{1}{n}\inf_{s_1\in \mathcal{S}_1}I(U;Y^n_{s_1})$\\
$A_{0_{\mathcal{S}_2}}=\frac{1}{n}\inf_{s_2\in \mathcal{S}_2}I(U;Y^n_{s_2})$\\ \label{1}
\item $A_{0_{\mathcal{S}_1}}=\frac{1}{n}\inf_{s_1\in \mathcal{S}_1}I(U;Z^n_{s_1})$\\
$A_{0_{\mathcal{S}_2}}=\frac{1}{n}\inf_{s_2\in \mathcal{S}_2}I(U;Z^n_{s_2})$\\ \label{2}
\item $A_{0_{\mathcal{S}_1}}=\frac{1}{n}\inf_{s_1\in \mathcal{S}_1}I(U;Y^n_{s_1})$\\
$A_{0_{\mathcal{S}_2}}=\frac{1}{n}\inf_{s_2\in \mathcal{S}_2}I(U;Z^n_{s_2})$\\ \label{3}
\item $A_{0_{\mathcal{S}_1}}=\frac{1}{n}\inf_{s_1\in \mathcal{S}_1}I(U;Z^n_{s_1})$\\
$A_{0_{\mathcal{S}_2}}=\frac{1}{n}\inf_{s_2\in \mathcal{S}_2}I(U;Y^n_{s_2})$\\ \label{4} 
\end{enumerate}

For Case \ref{1}, we have 
\begin{align}
&\Big|A_{0_{\mathcal{S}_1}}-A_{0_{\mathcal{S}_2}}\Big|\nonumber\\
&\quad\quad= \Big| \frac{1}{n}\inf_{s_1\in \mathcal{S}_1}I(U;Y^n_{s_1})-\frac{1}{n}\inf_{s_2\in \mathcal{S}_2}I(U;Y^n_{s_2})\Big|.\label{diff1}
\end{align}
Let $\eta>0$ be arbitrary. There exists an $\hat{s}_1=\hat{s}_1(\eta)$ such that
\begin{equation}\label{s1}
\inf_{s_1\in\mathcal{S}_1} I(U;Y^n_{s_1})\geq I(U;Y^n_{\hat{s}_1})-\eta.
\end{equation}
Since $D(\mathfrak{W}_1,\mathfrak{W}_2)<\epsilon$, there is an $\hat{s}_2=\hat{s}_2(\hat{s}_1)$ such that
\begin{equation}\label{s2}
d(W_{\hat{s}_1},W_{\hat{s}_2})<\epsilon.
\end{equation}
We can now apply Lemma \ref{lem2} (We let $U$ in \eqref{delta2} be a constant and we let $U$ in \eqref{diff1} take the role of $V$ in \eqref{delta2}). By \eqref{s2}, we have 
\begin{equation}\label{infs1}
\Big|I(U;Y^n_{\hat{s}_1})-I(U;Y^n_{\hat{s}_2})\Big|\leq n\delta_2(\epsilon,|\mathcal{Y}|).
\end{equation}
Combining \eqref{s1} and \eqref{infs1} we obtain
\begin{align*}
\inf_{s_1\in\mathcal{S}_1} I(U;Y^n_{s_1})&\geq I(U;Y^n_{\hat{s}_2})-n\delta(\epsilon,|\mathcal{Y}|)-\eta\\
&\geq \inf_{s_2\in\mathcal{S}_2}I(U;Y^n_{s_2})-n\delta_2(\epsilon,|\mathcal{Y}|)-\eta.
\end{align*}
This inequality holds for all $\eta>0$, we then obtain
\begin{align*}
\inf_{s_1\in\mathcal{S}_1} I(U;Y^n_{s_1})> \inf_{s_2\in\mathcal{S}_2}I(U;Y^n_{s_2})-n\delta_2(\epsilon,|\mathcal{Y}|).
\end{align*}
By changing the roles of $\mathcal{S}_1$ and $\mathcal{S}_2$ in the previous derivation, we get
\begin{equation*}
\Big|\inf_{s_1\in\mathcal{S}_1} I(U;Y^n_{s_1})-\inf_{s_2\in\mathcal{S}_2} I(U;Y^n_{s_2})\Big|\leq n\delta_2(\epsilon,|\mathcal{Y}|).
\end{equation*}
Using the same line of arguments as for Case \ref{1}, for Case \ref{2}, we have
\begin{equation*}
\Big|\inf_{s_1\in\mathcal{S}_1} I(U;Z^n_{s_1})-\inf_{s_2\in\mathcal{S}_2} I(U;Z^n_{s_2})\Big|\leq n\delta_2(\epsilon,|\mathcal{Z}|)
\end{equation*}
In Case \ref{3} and Case \ref{4} we have that for one compound BCC the maximum achievable common rate depends on the random variable $Y$ and for the other, the maximum achievable common rate depends on the random variable $Z$. We first study Case \ref{3}. We have
\begin{align*}
B_{0_{\mathcal{S}_1}}=\frac{1}{n}\inf_{s_1\in\mathcal{S}_1} I(U;Z^n_{s_1})&\geq \frac{1}{n}\inf_{s_1\in\mathcal{S}_1} I(U;Y^n_{s_1})=A_{0_{\mathcal{S}_1}}\\
B_{0_{\mathcal{S}_2}}=\frac{1}{n}\inf_{s_2\in\mathcal{S}_2} I(U;Y^n_{s_2})&\geq \frac{1}{n}\inf_{s_2\in\mathcal{S}_2} I(U;Z^n_{s_2})=A_{0_{\mathcal{S}_2}}.
\end{align*}
We have six possibilities to relate the two previous inequalities:
\begin{enumerate}
\renewcommand{\labelenumi}{\textbf{\theenumi}}
\renewcommand{\theenumi}{\Roman{enumi})}
\item $B_{0_{\mathcal{S}_1}}\geq A_{0_{\mathcal{S}_1}}\geq B_{0_{\mathcal{S}_2}}\geq A_{0_{\mathcal{S}_2}}$ and Lemma \ref{lem2} implies
\begin{align*}
\Big|A_{0_{\mathcal{S}_1}}-A_{0_{\mathcal{S}_2}}\Big|
&\leq \Big|B_{0_{\mathcal{S}_1}}-A_{0_{\mathcal{S}_2}}\Big|\leq \delta_2(\epsilon,|\mathcal{Z}|)
\end{align*}
\item $B_{0_{\mathcal{S}_1}}\geq B_{0_{\mathcal{S}_2}}\geq A_{0_{\mathcal{S}_1}}\geq A_{0_{\mathcal{S}_2}}$ implying
\begin{align*}
|A_{0_{\mathcal{S}_1}}-A_{0_{\mathcal{S}_2}}|\leq|B_{0_{\mathcal{S}_1}}-A_{0_{\mathcal{S}_2}}| \leq \delta_2(\epsilon,|\mathcal{Z}|)
\end{align*}
\item $B_{0_{\mathcal{S}_1}}\geq B_{0_{\mathcal{S}_2}}\geq A_{0_{\mathcal{S}_2}}\geq A_{0_{\mathcal{S}_1}}$ implying
\begin{align*}
|A_{0_{\mathcal{S}_1}}-A_{0_{\mathcal{S}_2}}|\leq | A_{0_{\mathcal{S}_1}}-B_{0_{\mathcal{S}_2}}|\leq \delta_2(\epsilon,|\mathcal{Y}|)
\end{align*}
\item $ B_{0_{\mathcal{S}_2}}\geq A_{0_{\mathcal{S}_2}}\geq B_{0_{\mathcal{S}_1}}\geq A_{0_{\mathcal{S}_1}}$ implying
\begin{align*}
|A_{0_{\mathcal{S}_1}}-A_{0_{\mathcal{S}_2}}|\leq | A_{0_{\mathcal{S}_1}}-B_{0_{\mathcal{S}_2}}|\leq \delta_2(\epsilon,|\mathcal{Y}|)
\end{align*}
\item $ B_{0_{\mathcal{S}_2}}\geq B_{0_{\mathcal{S}_1}}\geq A_{0_{\mathcal{S}_2}}\geq A_{0_{\mathcal{S}_1}}$ implying
\begin{align*}
|A_{0_{\mathcal{S}_1}}-A_{0_{\mathcal{S}_2}}|\leq | A_{0_{\mathcal{S}_1}}-B_{0_{\mathcal{S}_2}}|\leq \delta_2(\epsilon,|\mathcal{Y}|)
\end{align*}
\item $ B_{0_{\mathcal{S}_2}}\geq B_{0_{\mathcal{S}_1}}\geq A_{0_{\mathcal{S}_1}}\geq A_{0_{\mathcal{S}_2}}$ implying
\begin{align*}
|A_{0_{\mathcal{S}_1}}-A_{0_{\mathcal{S}_2}}|\leq |A_{0_{\mathcal{S}_2}}-B_{0_{\mathcal{S}_1}}| \leq \delta_2(\epsilon,|\mathcal{Z}|)
\end{align*}
\end{enumerate}
We use the same line of arguments for Case \ref{4} as for Case \ref{3} to bound the distance between the two maximum achievable common rates. It then holds for all cases
\begin{align*}
|A_{0_{\mathcal{S}_1}}-A_{0_{\mathcal{S}_2}}|&\leq \max\{\delta_2(\epsilon,|\mathcal{Y}|),\delta_2(\epsilon,|\mathcal{Y}|)\}\\
&=4H_2(\epsilon)+4\epsilon\max\{\log{|\mathcal{Y}|},\log{|\mathcal{Z}|}\}.
\end{align*}

\subsubsection{Confidential Message Rate}
Using the same line of arguments as in Case \ref{1} for the common-message rate, we get
\begin{align*}
|A_{1_{\mathcal{S}_1}}-A_{1_{\mathcal{S}_2}}&|\!=\!\Big|\frac{1}{n}\inf_{s_1\in\mathcal{S}_1} I(V;Y^n_{s_1}|U)\!-\!\frac{1}{n}\sup_{s_1\in\mathcal{S}_1} I(V;Z^n_{s_1}|U)\\
&-\frac{1}{n}\inf_{s_2\in\mathcal{S}_2} I(V;Y^n_{s_2}|U)\!+\!\frac{1}{n}\sup_{s_2\in\mathcal{S}_2} I(V;Z^n_{s_2}|U)\Big|\\ 
&\leq\frac{1}{n}\Big|\inf_{s_1\in\mathcal{S}_1} I(V;Y^n_{s_1}|U)-\inf_{s_2\in\mathcal{S}_2} I(V;Y^n_{s_2}|U)\Big|\\
&+\frac{1}{n}\Big|\inf_{s_2\in\mathcal{S}_2} I(V;Z^n_{s_2}|U)-\inf_{s_1\in\mathcal{S}_1} I(V;Z^n_{s_1}|U)\Big|\\
&\leq \delta_2(\epsilon,|\mathcal{Y}|)+\delta_2(\epsilon,|\mathcal{Z}|)\\
&\leq 4\epsilon\log|\mathcal{Y}||\mathcal{Z}|+8H_2(\epsilon).
\end{align*}
%\begin{equation*}
%\Big|\inf_{s_1\in\mathcal{S}_1} I(V;Y^n_{s_1}|U)-\inf_{s_2\in\mathcal{S}_2} I(V;Y^n_{s_2}|U)\Big|\leq \delta_2(\epsilon,|\mathcal{Y}|)
%\end{equation*}
%and 
%\begin{equation*}
%\Big|\sup_{s_1\in\mathcal{S}_1} I(V;Z^n_{s_1}|U)-\sup_{s_2\in\mathcal{S}_2} I(V;Z^n_{s_2}|U)\Big|\leq 4n(\epsilon\log{|\mathcal{Z}|}+H_2(\epsilon))
%\end{equation*}
%\begin{align*}
%\Big|\inf_{s_1\in\mathcal{S}_1} I(V;Y^n_{s_1}|U)-\sup_{s_1\in\mathcal{S}_1} I(V;Z^n_{s_1}|U)&-(\inf_{s_2\in\mathcal{S}_2} I(V;Y^n_{s_2}|U)-\sup_{s_2\in\mathcal{S}_2} I(V;Z^n_{s_2}|U))\Big|\\
%&\leq\Big|\inf_{s_1\in\mathcal{S}_1} I(V;Y^n_{s_1}|U)-\inf_{s_2\in\mathcal{S}_2} I(V;Y^n_{s_2}|U)\Big|-\Big|(\sup_{s_1\in\mathcal{S}_1} I(V;Z^n_{s_1}|U)-\sup_{s_2\in\mathcal{S}_2} I(V;Z^n_{s_2}|U))\Big|\\
%&\leq n\delta(\epsilon,\log|\mathcal{Y}|)+n\delta(\epsilon,\log|\mathcal{Z}|)
%\end{align*}
%This implies for the maximum achievable confidential rate
%\begin{align*}
%\frac{1}{n}\Big|\inf_{s_1\in\mathcal{S}_1} I(V;Y^n_{s_1}|U)-\sup_{s_1\in\mathcal{S}_1} I(V;Z^n_{s_1}|U)\Big|&-%\frac{1}{n}\Big|\inf_{s_2\in\mathcal{S}_2} I(V;Y^n_{s_2}|U)-\sup_{s_2\in\mathcal{S}_2} I(V;Z^n_{s_2}|U)\Big|\\
%&=\Big|\frac{1}{n}\inf_{s_1\in\mathcal{S}_1} I(V;Y^n_{s_1}|U)-\frac{1}{n}\sup_{s_1\in\mathcal{S}_1} I(V;Z^n_{s_1}|U)\Big|\\
%&\;\;+\Big|\frac{1}{n}\inf_{s_2\in\mathcal{S}_2} I(V;Y^n_{s_2}|U)-\frac{1}{n}\sup_{s_2\in\mathcal{S}_2} I(V;Z^n_%{s_2}|U)\Big|\\
%&\leq \delta_2(\epsilon,|\mathcal{Y}|)+\delta_2(\epsilon,|\mathcal{Z}|)
%\end{align*}
\end{IEEEproof}

\begin{thm}\label{capacitycont}
Let $\epsilon \in (0,1)$. Let $\mathfrak{W}_1$ and $\mathfrak{W}_2$ be two compound BCCs. If
\begin{equation}\label{difchan}
D(\mathfrak{W}_1,\mathfrak{W}_2)\leq \epsilon
\end{equation}
then it holds

\begin{equation*}
D_R(\mathcal{C}(\mathfrak{W}_1),\mathcal{C}(\mathfrak{W}_2))\leq \delta(\epsilon,|\mathcal{Y}|,|\mathcal{Z}|).
\end{equation*}
\end{thm}
\begin{IEEEproof}
We define the sets $\mathcal{D}_1,\mathcal{B}_1\subset\mathbb{R}^2_+$   and 
\begin{align*}
\mathcal{D}_1&=\bigcup_{n\in\mathbb{N}}\bigcup_{U-V-X^n}\mathcal{R}_n(\mathfrak{W}_1,U,V,X^n)\\
\mathcal{B}_1&=\mathcal{C}(\mathfrak{W}_1)\backslash\bigcup_{n\in\mathbb{N}}\bigcup_{U-V-X^n}\mathcal{R}_n(\mathfrak{W}_1,U,V,X^n)
\end{align*}
with random variables $U-V-X^n$ forming a Markov chain. 
Let $(R_{0_{\mathcal{S}_1}},R_{1_{\mathcal{S}_1}})\in\mathcal{D}_1$. Then there exists a $n\in\mathbb{N}$ and random variables satisfying the Markov chain relationship $\hat{U}-\hat{V}-\hat{X^n}$ such that $(R_{0_{\mathcal{S}_1}},R_{1_{\mathcal{S}_1}})\in\mathcal{R}_n(\mathfrak{W}_1,\hat{U},\hat{V},\hat{X^n})$. From Lemma \ref{ratecont} and \eqref{difchan} we have that
\begin{equation*}
d(\mathcal{R}_n(\mathfrak{W}_1,\hat{U},\hat{V},\hat{X^n}),\mathcal{R}_n(\mathfrak{W}_2,\hat{U},\hat{V},\hat{X^n}))\leq \delta(\epsilon,|\mathcal{Y}||\mathcal{Z}|).
\end{equation*}
 This means that there exists a rate pair $(R_{0_{\mathcal{S}_2}}(R_{0_{\mathcal{S}_1}}),R_{1_{\mathcal{S}_2}}(R_{1_{\mathcal{S}_1}}))\in\mathcal{R}_n(\mathfrak{W}_2,\hat{U},\hat{V},\hat{X^n})$ such that
\begin{equation*}
{|R_{0_{\mathcal{S}_1}}-R_{0_{\mathcal{S}_2}}|+|R_{1_{\mathcal{S}_1}}-R_{1_{\mathcal{S}_2}}|}\leq \delta(\epsilon,|\mathcal{Y}|,|\mathcal{Z}|).
\end{equation*}
Let $(\hat{R}_{0_{\mathcal{S}_1}},\hat{R}_{1_{\mathcal{S}_1}})\in\mathcal{B}_1$. Then there exist two rate pairs $(\dot{R}_{0_{\mathcal{S}_1}},\dot{R}_{1_{\mathcal{S}_1}}),(\tilde{R}_{0_{\mathcal{S}_1}},\tilde{R}_{1_{\mathcal{S}_1}}) \in\mathcal{D}_1$ such that
\begin{align*}
\hat{R}_{0_{\mathcal{S}_1}}&=\lambda \dot{R}_{0_{\mathcal{S}_1}} +(1-\lambda)\tilde{R}_{0_{\mathcal{S}_1}}\\
\hat{R}_{1_{\mathcal{S}_1}}&=\lambda \dot{R}_{1_{\mathcal{S}_1}} +(1-\lambda)\tilde{R}_{1_{\mathcal{S}_1}}
\end{align*}
for some $\lambda\in(0,1)$. For each  $(\dot{R}_{0_{\mathcal{S}_1}},\dot{R}_{1_{\mathcal{S}_1}})$ and $(\tilde{R}_{0_{\mathcal{S}_1}},\tilde{R}_{1_{\mathcal{S}_1}})$ there exist random variables satisfying the Markov chain relation $\dot{U}-\dot{V}-\dot{X}^n$ and $\tilde{U}-\tilde{V}-\tilde{X^n}$ such that $(\dot{R}_{0_{\mathcal{S}_1}},\dot{R}_{1_{\mathcal{S}_1}})\in\mathcal{R}_n(\mathfrak{W}_1,\dot{U},\dot{V},\dot{X}^n)$ and $(\tilde{R}_{0_{\mathcal{S}_1}},\tilde{R}_{1_{\mathcal{S}_1}})\in\mathcal{R}_n(\mathfrak{W}_1,\tilde{U},\tilde{V},\tilde{X}^n)$. Then from Lemma \ref{ratecont} and \eqref{difchan} we have that there exist rate pairs $(\dot{R}_{0_{\mathcal{S}_2}}(\dot{R}_{0_{\mathcal{S}_1}}),\dot{R}_{1_{\mathcal{S}_2}}(\dot{R}_{1_{\mathcal{S}_1}}))\in\mathcal{R}_n(\mathfrak{W}_2,\dot{U},\dot{V},\dot{X}^n)$ and $(\tilde{R}_{0_{\mathcal{S}_2}}(\tilde{R}_{0_{\mathcal{S}_1}}),\tilde{R}_{1_{\mathcal{S}_2}}(\tilde{R}_{1_{\mathcal{S}_1}}))\in\mathcal{R}_n(\mathfrak{W}_2,\tilde{U},\tilde{V},\tilde{X}^n)$ such that
\begin{align*}
{|\dot{R}_{0_{\mathcal{S}_1}}-\dot{R}_{0_{\mathcal{S}_2}}|+|\dot{R}_{1_{\mathcal{S}_1}}-\dot{R}_{1_{\mathcal{S}_2}}|}&\leq \delta(\epsilon,|\mathcal{Y}|,|\mathcal{Z}|)\\
{|\tilde{R}_{0_{\mathcal{S}_1}}-\tilde{R}_{0_{\mathcal{S}_2}}|+|\tilde{R}_{1_{\mathcal{S}_1}}-\tilde{R}_{1_{\mathcal{S}_2}}|}&\leq \delta(\epsilon,|\mathcal{Y}|,|\mathcal{Z}|).
\end{align*}
Then there is a rate pair $(\hat{R}_{0_{\mathcal{S}_2}},\hat{R}_{1_{\mathcal{S}_2}})\in\mathcal{C}(\mathfrak{W}_2)$ with
\begin{align*}
\hat{R}_{0_{\mathcal{S}_2}}&=\lambda \dot{R}_{0_{\mathcal{S}_2}} +(1-\lambda)\tilde{R}_{0_{\mathcal{S}_2}}\\
\hat{R}_{1_{\mathcal{S}_2}}&=\lambda \dot{R}_{1_{\mathcal{S}_2}} +(1-\lambda)\tilde{R}_{1_{\mathcal{S}_2}}.
\end{align*}
Further we have
\begin{align*}
|\hat{R}_{0_{\mathcal{S}_1}}-\hat{R}_{0_{\mathcal{S}_2}}|&= |\lambda\dot{R}_{0_{\mathcal{S}_2}}+(1-\lambda)\tilde{R}_{0_{\mathcal{S}_2}}\\
&\;\;\;\;-\lambda \dot{R}_{0_{\mathcal{S}_1}}+(1-\lambda)\tilde{R}_{0_{\mathcal{S}_1}}|\\
&\leq \lambda |\dot{R}_{0_{\mathcal{S}_1}}-\dot{R}_{0_{\mathcal{S}_2}}| + (1-\lambda)|\tilde{R}_{0_{\mathcal{S}_1}}-\tilde{R}_{0_{\mathcal{S}_2}}|\\
&\leq \delta'(\epsilon,|\mathcal{Y}|,|\mathcal{Z}|)
\end{align*}
and using the same line of arguments
\begin{equation*}
|\hat{R}_{1_{\mathcal{S}_1}}-\hat{R}_{1_{\mathcal{S}_2}}|\leq \delta''(\epsilon,|\mathcal{Y}|,|\mathcal{Z}|).
\end{equation*}
This leads us to the following result
\begin{equation*}
{|\hat{R}_{0_{\mathcal{S}_1}}-\hat{R}_{0_{\mathcal{S}_2}}|+ |\hat{R}_{1_{\mathcal{S}_1}}-\hat{R}_{1_{\mathcal{S}_2}}|}\leq \delta(\epsilon,|\mathcal{Y}|,|\mathcal{Z}|).
\end{equation*}
We can conclude that for every rate pair $(R_{0_{\mathcal{S}_1}},R_{1_{\mathcal{S}_1}})\in\mathcal{C}(\mathfrak{W}_1)$ we can find a rate pair $(R_{0_{\mathcal{S}_2}}(R_{0_{\mathcal{S}_1}}),R_{1_{\mathcal{S}_2}}(R_{1_{\mathcal{S}_1}}))\in\mathcal{C}(\mathfrak{W}_2)$ such that
\begin{equation}\label{dichteq}
{|{R}_{0_{\mathcal{S}_1}}-{R}_{0_{\mathcal{S}_2}}|+ |{R}_{1_{\mathcal{S}_1}}-{R}_{1_{\mathcal{S}_2}}|}\leq \delta(\epsilon,|\mathcal{Y}|,|\mathcal{Z}|)
\end{equation}
We use the same line of arguments to show that for every rate pair $(R_{0_{\mathcal{S}_2}},R_{1_{\mathcal{S}_2}})\in\mathcal{C}(\mathfrak{W}_2)$ there is a rate pair $(R_{0_{\mathcal{S}_1}}(R_{0_{\mathcal{S}_2}}),R_{1_{\mathcal{S}_1}}(R_{1_{\mathcal{S}_2}}))\in\mathcal{C}(\mathfrak{W}_1)$ such that \eqref{dichteq} holds. This completes the proof.
\end{IEEEproof}

% ========================================================================================
% ========================================================================================
% ========================================================================================

\section{Discussion}
\label{sec4}
This work was motivated by the question whether the compound BCC capacity region depends continuously on the uncertainty set or not. We have shown that the compound BCC model is robust, i.e., small changes in the uncertainty set lead to small changes in the capacity region, which is desirable.

Let's see what happens when the user's CSI is reduced further. For example, the AVBCC is described by the same uncertainty set as the compound BCC, but in addition, the actual channel realization varies from channel use to channel use in an arbitrary fashion. The AVBCC can be used for example to model the presence of jamming, see \cite{boche2014continuity}. This may lead the channel to "emulate" a valid input, impeding the legitimate receiver to decide on the correct codeword. This property is known as symmetrizability; see \cite[Sec.~III, Def.~5]{boche2014continuity}

We adapt the AVC example from \cite[Sec.~V]{boche2014continuity} to the channel of receiver 1 of the AVBCC, where the input and the output alphabets are of size $|\mathcal{X}|=2$ and $|\mathcal{Y}|=3$, respectively, and the uncertainty set consists of only two elements, i.e., $|\mathcal{S}|=2$. The AVC to receiver 1 is given by $\mathcal{W}(\lambda)=\{W_1(\lambda),W_2(\lambda)\}$  with 
\begin{equation*}
	W_1(\lambda)=\begin{pmatrix}1 & 0 & 0 \\ 1 & \lambda & 1-\lambda \end{pmatrix} \quad\!\!\!\!\text{and}\!\!\!\!\quad
	W_2(\lambda)=\begin{pmatrix}\lambda & 0 & 1-\lambda \\ 0 & 1 & 0 \end{pmatrix}
\end{equation*}
where $\lambda \in [0,1]$. The AVC $\mathcal{V}$ to receiver 2 has an output alphabet of size $|\mathcal{Z}|=2$ and is defined as $\mathcal{V}=\{V,V\}$ with
\begin{equation*}
	V=\begin{pmatrix}\frac{1}{2} & \frac{1}{2} \\[0.1 cm] \frac{1}{2} & \frac{1}{2} \end{pmatrix}.
\end{equation*}
In \cite[Sec.~V]{boche2014continuity}, it is shown that the AVC $\mathcal{W}(\lambda)$ is non-symmetrizable for all $\lambda \in (0,1]$, and symmetrizable for $\lambda=0$, in which case the capacity region collapses to the point $(0,0)\in\mathbb{R}^2_+$. Following the argumentation in \cite[Sec.~V]{boche2014continuity}, it can be shown that capacity region is indeed discontinuous in $\lambda=0$.

\bibliographystyle{IEEEtran}
\bibliography{IEEEabrv,confs-jrnls,agrigorebib}

\newpage
\onecolumn

\appendix

Here we present the proof of Lemma \ref{lem2} based on \cite{boche2014continuity}.
\begin{proof}
Let $0\leq k\leq n$ be arbitrary. We define
\begin{equation*}
P_{UVY_1^k\widetilde{Y}_{k+1}^n}(u,v,y^k_1,{y}^n_{k+1}):=\sum_{x^n\in\mathcal{X}^n}\prod_{l=1}^k W(y_l|x_l)\prod_{l=k+1}^n \widetilde{W}(y_l|x_l)E(x^n|v)P_{V|U}(v|u)P_{U}(u).
\end{equation*}
So we have
\begin{equation}
I(V;Y^n|U)-I(V;\widetilde{Y}^n|U)=\sum_{k=0}^{n-1}\Big(I(V;Y_1^{k+1}\widetilde{Y}_{k+2}^n|U)-I(V;Y^k_1\widetilde{Y}_{k+1}^n|U)\Big).
\end{equation}
For all $0\leq k\leq n-1$ it holds
\begin{align}
I(V;Y^{k+1}_1\widetilde{Y}_{k+2}^n|U)-I(V;Y^k_1\widetilde{Y}_{k+1}^n|U)&= I(V;Y^k_1|U)+I(V;Y_{k+1}\widetilde{Y}_{k+2}^n|Y^k_1U)-I(V;Y^k_1|U)-I(V;\widetilde{Y}_{k+1}^n|Y^k_1U)\nonumber\\&= I(V;Y_{k+1}\widetilde{Y}_{k+2}^n|Y^k_1U)-I(V;\widetilde{Y}_{k+1}^n|Y^k_1U)\nonumber\\
&= I(V;\widetilde{Y}_{k+2}^n|Y^k_1U)+I(V;Y_{k+1}|\widetilde{Y}_{k+2}^nY^k_1U)\nonumber\\
&\;\;\;\;\;-I(V;\widetilde{Y}_{k+2}^n|Y^k_1U)-I(V;\widetilde{Y}_{k+1}|\widetilde{Y}_{k+2}^nY^k_1U)\nonumber\\
&= I(V;Y_{k+1}|\widetilde{Y}_{k+2}^nY^k_1U)-I(V;\widetilde{Y}_{k+1}|\widetilde{Y}_{k+2}^nY^k_1U)\nonumber\\
&=H(Y_{k+1}|\widetilde{Y}_{k+2}^nY^k_1U)-H(\widetilde{Y}_{k+1}|\widetilde{Y}_{k+2}^nY^k_1U)\nonumber\\
&\;\;\;\;\;-H(VY_{k+1}|\widetilde{Y}_{k+2}^nY^k_1U)+H(V\widetilde{Y}_{k+1}|\widetilde{Y}_{k+2}^nY^k_1U).\label{eqlem}
\end{align}
We want to analyze the right-hand side of \eqref{eqlem}. For $0\leq k\leq n-1$, it holds
\begin{align*}
\|P_{UVY^{k+1}_1\widetilde{Y}^{n}_{k+2}}-P_{UVY^{k}_1\widetilde{Y}^{n}_{k+1}}\|&=\sum_{v\in\mathcal{V}}\sum_{u\in\mathcal{U}}\sum_{y^n\in\mathcal{Y}^n}\Big|P_{UVY^{k+1}_1\widetilde{Y}^{n}_{k+2}}(u,v,y^{k+1}_1y^{n}_{k+2})-P_{UVY^{k}_1\widetilde{Y}^{n}_{k+1}}(u,v,y^{k}_1y^{n}_{k+1})\Big|\\
&=\sum_{v\in\mathcal{V}}\sum_{u\in\mathcal{U}}\sum_{y^n\in\mathcal{Y}^n}\Big|\sum_{x^n\in\mathcal{X}^n}\Big(\prod_{l=1}^{k+1} W(y_l|x_l)\prod_{l=k+2}^n \widetilde{W}(y_l|x_l)-\prod_{l=1}^{k+1} W(y_l|x_l)\prod_{l=k+2}^n \widetilde{W}(y_l|x_l)\Big)\\
&\;\;\;\;\;\times E(x^n|v)P_{V|U}(v|u)P_{U}(u)\Big|\\
&=\sum_{v\in\mathcal{V}}\sum_{u\in\mathcal{U}}\sum_{y^n\in\mathcal{Y}^n}\Big|\sum_{x^n\in\mathcal{X}^n}\prod_{l=1}^{k} W(y_l|x_l)\prod_{l=k+2}^n \widetilde{W}(y_l|x_l)\Big(W(y_{k+1}|x_{k+1})-\widetilde{W}(y_{k+1}|x_{k+1})\Big)\\
&\;\;\;\;\;\times E(x^n|v)P_{V|U}(v|u)P_{U}(u)\Big|\\
&\leq\sum_{v\in\mathcal{V}}\sum_{u\in\mathcal{U}}\sum_{y^n\in\mathcal{Y}^n}\sum_{x^n\in\mathcal{X}^n}\prod_{l=1}^{k} W(y_l|x_l)\prod_{l=k+2}^n \widetilde{W}(y_l|x_l)\Big|W(y_{k+1}|x_{k+1})-\widetilde{W}(y_{k+1}|x_{k+1})\Big|\\
&\;\;\;\;\;\times E(x^n|v)P_{V|U}(v|u)P_{U}(u)\\
&=\sum_{v\in\mathcal{V}}\sum_{u\in\mathcal{U}}\sum_{x^n\in\mathcal{X}^n}\Big(\sum_{y^n\in\mathcal{Y}^n}\prod_{l=1}^{k} W(y_l|x_l)\prod_{l=k+2}^n \widetilde{W}(y_l|x_l)\Big|W(y_{k+1}|x_{k+1})-\widetilde{W}(y_{k+1}|x_{k+1})\Big|\Big)\\
&\;\;\;\;\;\times E(x^n|v)P_{V|U}(v|u)P_{U}(u)\\
&=\sum_{u\in\mathcal{U}}\sum_{x^n\in\mathcal{X}^n}\sum_{y_{k+1}\in\mathcal{Y}}\Big|W(y_{k+1}|x_{k+1})-\widetilde{W}(y_{k+1}|x_{k+1})\Big|\\
&\;\;\;\;\;\times E(x^n|v)P_{V|U}(v|u)P_{U}(u)\\
&<\epsilon\sum_{v\in\mathcal{V}}\sum_{u\in\mathcal{U}}\sum_{x^n\in\mathcal{X}^n}E(x^n|v)P_{V|U}(v|u)P_{U}(u)=\epsilon.
\end{align*}
Which shows that the total variation between the joint probability distribution $P_{UVY^k\widetilde{Y}_{k+1}^n}$ and $P_{UVY^{k+1}\widetilde{Y}_{k+2}^n}$ is smaller than $\epsilon$. Then by Lemma \ref{lem1} it holds
\begin{equation}\label{eq1lem}
\Big|H(Y_{k+1}|\widetilde{Y}_{k+2}^nY^k_1U)-H(\widetilde{Y}_{k+1}|\widetilde{Y}_{k+2}^nY^k_1U)\Big|<2\epsilon\log|\mathcal{Y}|+2H_2(\epsilon)
\end{equation}
and
\begin{align}
\Big|H(VY_{k+1}|\widetilde{Y}_{k+2}^nY^k_1U)-H(V\widetilde{Y}_{k+1}|\widetilde{Y}_{k+2}^nY^k_1U)\Big|&=\Big|H(V|\widetilde{Y}_{k+2}^nY^k_1U)+H(Y_{k+1}|V\widetilde{Y}_{k+2}^nY^k_1U) \nonumber\\ 
&\;\;\;\;-H(V|\widetilde{Y}_{k+2}^nY^k_1U)-H(\widetilde{Y}_{k+1}|V\widetilde{Y}_{k+2}^nY^k_1U)\Big| \nonumber\\
&=\Big|H(Y_{k+1}|V\widetilde{Y}_{k+2}^nY^k_1U)-H(\widetilde{Y}_{k+1}|V\widetilde{Y}_{k+2}^nY^k_1U)\Big|\nonumber\\ 
&<2\epsilon\log|\mathcal{Y}|+2H_2(\epsilon) \label{eq2lem}
\end{align}
Inserting \eqref{eq1lem} and \eqref{eq2lem} into \eqref{eqlem} we obtain
\begin{equation}
\Big|I(V;Y^{k+1}_1\widetilde{Y}_{k+2}^n|U)-I(V;Y^k_1\widetilde{Y}_{k+1}^n|U)\Big|\leq4\epsilon\log|\mathcal{Y}|+4H_2(\epsilon)=:\delta_2(\epsilon,|\mathcal{Y}|) 
\end{equation}
This gives in particular the following upper bound for the difference between $I(V;Y^n|U)$ and $I(V;\widetilde{Y}^n|U)$
\begin{align*}
\Big|I(V;Y^n|U)-I(V;\widetilde{Y}^n|U)\Big|&\leq \sum_{k=0}^{n-1}\Big|I(V;Y_1^{k+1}\widetilde{Y}_{k+2}^n|U)-I(V;Y^k_1\widetilde{Y}_{k+1}^n|U)\Big|\\
&\leq n\delta_2(\epsilon,|\mathcal{Y}|) 
\end{align*}
proving the lemma.
\end{proof}
\end{document}